\documentclass[aps,prc,groupedaddress,twocolumn]{revtex4-1}
\usepackage{graphicx}
\usepackage[breaklinks]{hyperref}
\usepackage[english]{babel}
\usepackage{booktabs}
\usepackage{rotating}
\usepackage{multirow}
\usepackage{amsmath}

\begin{document}

\title{Electrovolt-scale backgrounds from surfaces}

\author{Alan E. Robinson}
\affiliation{D\'epartement de physique, Universit\'e de Montr\'eal, Montr\'eal, Canada H3C 3J7}
\email{alan.robinson@umontreal.ca}

\date{October 21, 2020}

\begin{abstract}
Recent results from the SENSEI experiment show that a cut on event clustering can reduce low-energy excesses in their eV-sensitive calorimeter.  This hints at the role of surrounding uninstrumented surfaces in producing backgrounds.  Charged particles crossing dielectric boundaries are well known to produce low-energy radiation.  In particular, transition radiation, secondary electron emission, and sputtering may contribute to the spectrum, morphology, and rate of events observed in eV-sensitive detectors.  The rich phenomenology and high yields of these surface processes will complicate comparisons of low-threshold dark matter detectors both to each other and to background models.
\end{abstract}


\maketitle

%

While pursing expanded sensitivity to dark matter scattering, the record low eV-scale thresholds achieved by SuperCDMS \cite{HVeV1}, SENSEI \cite{SENSEI1}, and EDELWEISS \cite{EDELWEISS-III} also observe rare low-energy effects in the scattering of matter and radiation.  All three experiments report an excess of events at low energy that is otherwise not explained by previously considered backgrounds in particle physics detectors.

Recently SENSEI, using CCDs with 15 $\mu$m pitch pixels, showed that their background of pixels containing 1 or 2 conduction electrons is significantly reduced when applying a spacial clustering cut to their data \cite{barak2020seisei}.  The scale of the observed clustering of 0.3 mm (20 pixels) is similar to the distance in vacuum between the detector and its copper housing \cite{wineandcheese}.  This indicates that these low-energy quanta are efficiently radiated from the copper surfaces surrounding the detector.  The rate of these low-charge hits is also spatially correlated with high-energy tracks that deposit more than 100 electrons ($>360$~eV) in the CCD.

The eV-energy quanta could be either photons, free electrons, or free atoms crossing the vacuum.  In copper and silicon, multi-keV charged radiation produces the former primarily through transition radiation \cite{10.2307/52971,[{For sub-keV incident radiation where transition radiation is suppressed, it has been shown in copper that interband transitions can produce some luminescence, but with an efficiency much smaller than 1, and thus negligible for this analysis.  }] PAPANICOLAOU1976403}, the second by secondary electron emission, and the latter by sputtering \cite{Eckstein2007}.  These processes, the secondary production yields and energies, and their detection in silicon and copper will be considered.  To match the SENSEI observations, the low-energy quanta must efficiently produced with secondary particle yields on order unity or greater in order to efficiently produce clusters.  The energy spectrum of the quanta must also be very soft, with a ratio of 5:1 or greater between the number of measured two-electron and three electron events.

\begin{figure*}
\def\svgwidth{0.96\columnwidth}
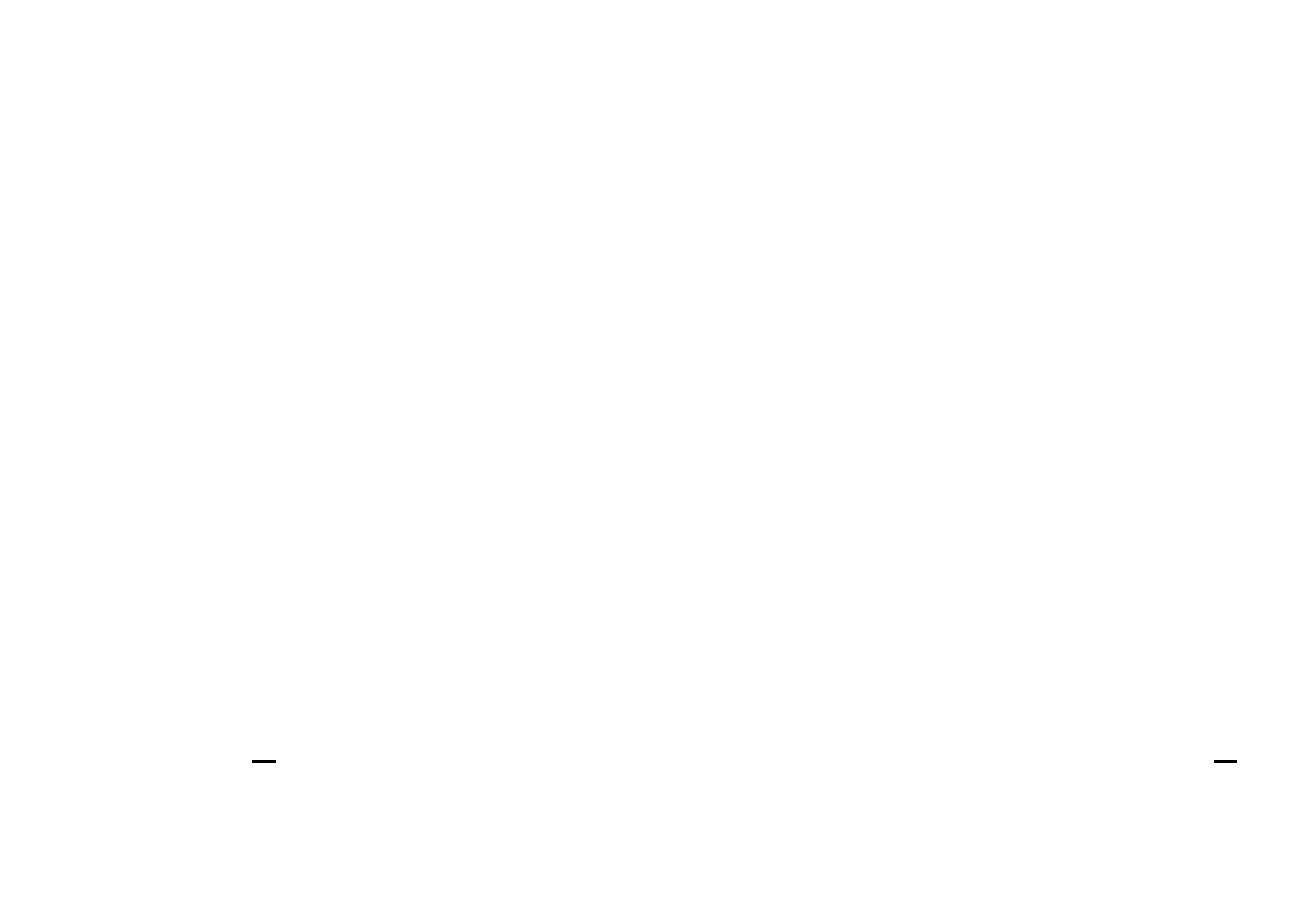
\def\svgwidth{0.96\columnwidth}
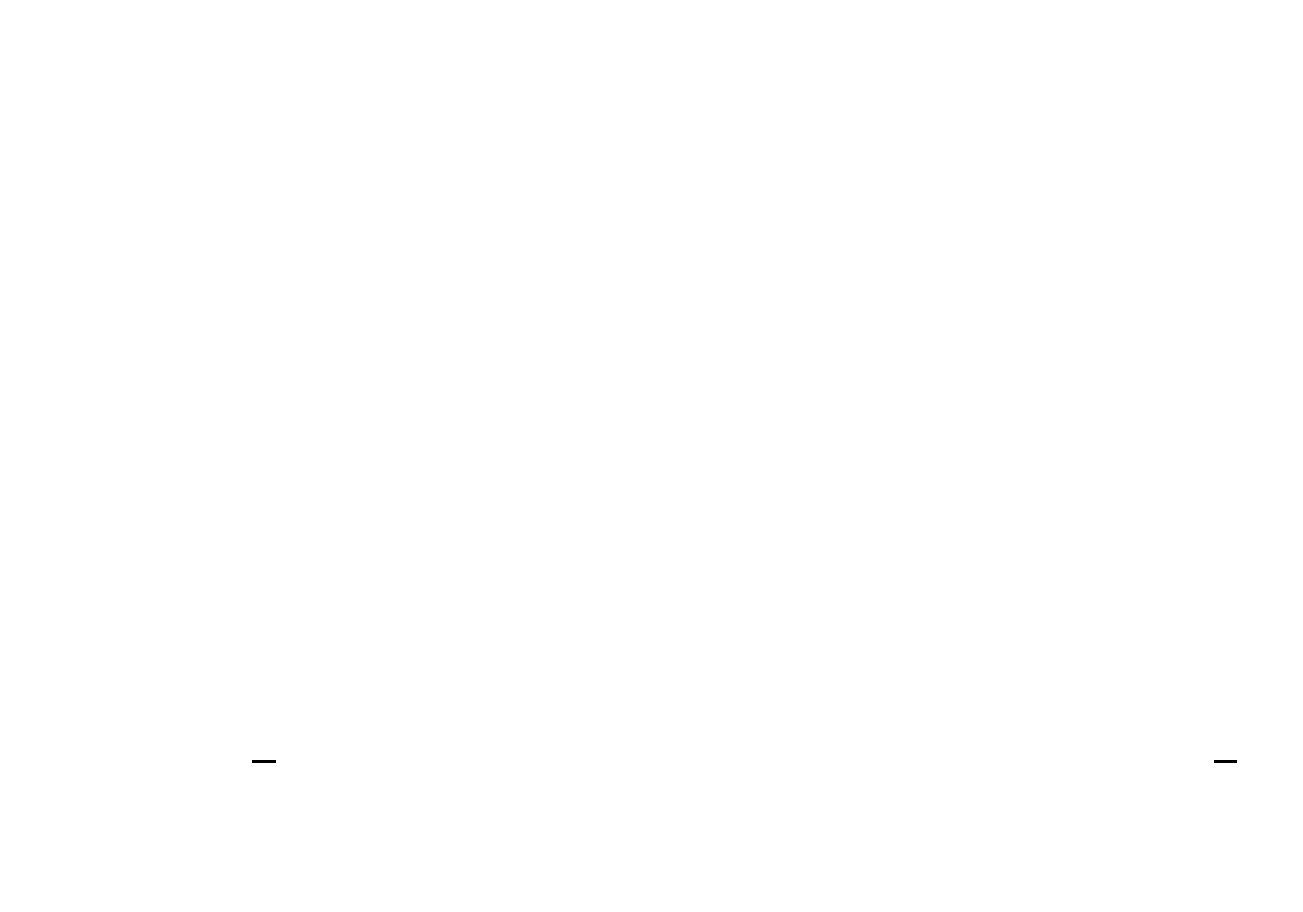
\caption{\label{fig:energy} Energy spectrum for photons produced by charged particles crossing between silicon and copper surfaces (left) and absorbed in silicon (right) for photons transiting both surfaces at normal incidence.  The most probable electron yield from \cite{Kurinsky} for a given photon energy is shown.  Using their model, a ratio of 0.6 3-electron events per 2-electron event in SENSEI is predicted.}
\end{figure*}

\begin{figure}
\def\svgwidth{0.96\columnwidth}
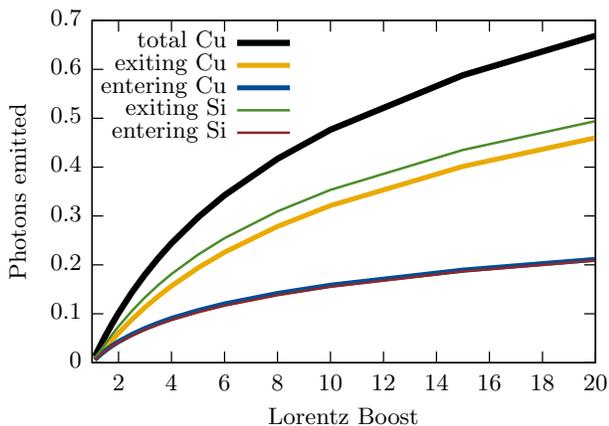
\caption{\label{fig:rate} Yield of transition radiation photons produced as a function of Lorentz boost of an electron or other singly charged particle crossing two vacuum gaps from copper to silicon to copper.  Most electrons from natural radioactivity are expected to be produced with Lorentz boosts less than 4.}
\end{figure}

Figure \ref{fig:energy} shows the expected angle-integrated photon energy distribution for charged particles transitioning normally through copper and silicon.  The significant rate of photons with energies greater than 8eV and the low yield of photons, shown in Figure \ref{fig:rate}, are inconsistent with the background observed by SENSEI. However, transition radiation is a easily modelled background that is expected to contribute subdominantly to the spectrum observed by SENSEI.  Notably, once transition radiation is produced, it can be very efficiently separated from the original charged particle track due to a large angular separation between the passing charged particle and the emitted photons.  The efficiency for reflecting the low-energy optical photons from copper surfaces is also very high, permitting an even wider spacial separation and motivating the need for tightly fitting hermetic detector housings and significant softening of the energy spectrum.

Secondary electrons, from ionization generated within $\mathcal{O}(10)$~\AA of a material surface, can very efficiently escape surfaces: it is one of the principle means of generating images in scanning electron microscopy.  The theory of secondary electron emission from copper and silicon has been well studied with \cite{10.2307/52971}.  Their kinetic energy spectra peak near 2.5~eV from copper surfaces and near 1.8~eV from silicon surfaces.  The lowest energies of these electrons can also be efficiently reflected from metallic surfaces when their momentum is below the Fermi momentum of the metal.  The secondary electron yield depends strongly on the near surface ionization produced, peaks for in-plane particle velocities near 0.03$c$, and varies mildly on the quality of the surface \cite[This latter effect has been well studied in order to suppress secondary electron emission from copper in high-luminosity accelerators.]{PhysRevSTAB.18.051002}.  Yields near 1--1.5 for normally incident primary electrons around 500~eV \cite{doi:10.1002/sca.20124} and yields of 5 for incident alpha radiation at 1.2 MeV have been measured \cite{PhysRevA.52.3959}.  These yields are sufficient to produce clustering as observed in SENSEI although the efficiency drops for higher velocity particles as a function of their electronic stopping power.  For minimum ionizing electrons, grazing trajectories, where the secondary electron yield is proportional to the secant of the incident angle, would be required to maintain the high yield required to observe clustering.

SENSEI is nominally insensitive to charge collected on surfaces.  The front and back sides of these CCDs are coated with either insulators or ohmic contacts that would transport and/or block electrons from reaching the active area of the detector.  In contrast, cryogenic bolometers or SuperCDMS would be sensitive to these secondary electrons, with the work function for silicon ($\sim$4.85~eV) added to the kinetic energy of the absorbed electrons.  Within SENSEI and other CCDs, secondary electrons could be observed if the electrons efficiently radiatively recombine, allowing the CCD to detect the emitted photons.  Recombination at trapped holes on Si-SiO$_2$ interfaces, either on native oxide surfaces or on buried layers, may be sufficiently efficient \cite{BARABAN2019102} to allow for clustering in SENSEI.  One potential means of identifying SiO$_2$ scintillation photons is by the ionization spectrum produced by the 4.2~eV they produce. Per \citet{Kurinsky}, a 7:1 ratio of 2-electron to 1-electron pixel charges would be expected from these quanta with no 3-electron pixels produced.

Sputtering by recoiling nuclei, notably by $^{206}$Pb from the decay of the radon daughter $^{210}$Po plated out on surfaces, could launch on the order of 50 recoiling atoms per decay from a copper surface towards a detector \cite{SARHOLTKRISTENSEN199185}.  While this yield may be efficient, these sputtered atoms would not penetrate sufficiently deeply into the silicon surface of SENSEI to produce ionization.  
In addition, the observed reduction of the background following the addition of external lead shielding in SENSEI is inconsistent with background mechanisms that are dependent on surface alpha activity.

While SENSEI operates with grounded detector and housing surfaces, other eV-sensitive detectors such as SuperCDMS apply a voltage bias between the detector and the detector housing.  This can accelerate sputtered ions or secondary electrons to energies where tertiary particle production is efficient, potentially resulting in significantly amplified energy spectra and sub-critical charge multiplication.  The observation of low-energy quanta in SuperCDMS HVeV detectors independent of detector bias seems to prefer models of backgrounds from neutral quanta, in particular transition radiation photons, despite their lower predicted yield as compared to secondary electrons.

In conclusion, high-energy charged particles can very efficiently produce low-energy quanta at detector surfaces, concentrating radioactive backgrounds in the energy range for which these detectors claim their greatest sensitivity to low-mass dark matter.  These low energy quanta are efficiently produced by multiple mechanisms, some of which are difficult to model, that need to all be considered for eV-sensitive detectors.  There remains several open questions as to the phenomenology of these backgrounds, in particular the role of cathodoluminescence and the efficiency of detecting surface deposited charge.  For the most easily modelled background, that from transition radiation, the production of optical photons by this mechanism is not currently implemented by Geant4 \cite{Geant4} or MCNP \cite{MCNP}, the radiation transport simulations used by SENSEI, SuperCDMS, and EDELWEISS.

Some strategies may efficiently mitigate eV-scale backgrounds, most notably the vetoing of high-energy radiation on surfaces surrounding the detector as is done for CRESST-III \cite{PhysRevD.100.102002}.  However, this experiment has not succeeded in eliminating its low-energy background by this means.  In addition, none of the mechanisms considered in this article fully explains the SENSEI background.  This does not alleviate the need to understand these backgrounds, but rather indicates that further study of low-energy quanta and surface physics is required to fully understand low background eV-sensitive detectors.

\begin{acknowledgments}
This research is supported by the Canada First Research Excellence Fund through the Arthur B. McDonald Canadian Astroparticle Physics Research Institute.
\end{acknowledgments}

\appendix*

\section{Calculating the transition radiation spectrum}

Transition radiation is a well known process that produces low-energy photons from the propagation of a charged particle across a dielectric discontinuity.  The fluence $W$, double differential in photon energy $\hbar\omega$ and angle $\theta$, of radiation emitted by a charged particle crossing a material / vacuum interface at normal incidence is \cite[][Eq 2.33]{GINZBURG19791}
\begin{widetext}
\begin{equation}
\frac{dW}{d\omega d\Theta} = \frac{4 z^2 \alpha \beta^2}{\pi}\cos^2\theta \sin^2\theta |1-\epsilon|^2 \times
\left| \frac{1 - \beta^2 \pm \beta \sqrt{\epsilon - \sin^2\theta}} {(1-\beta^2 \cos^2\theta)(1\pm\beta\sqrt{\epsilon-\sin^2\theta})(\epsilon\cos\theta+\sqrt{\epsilon-\sin^2\theta})} \right|^2
\end{equation}
\end{widetext}
where $z$ is the particle's charge, $\alpha$ is the fine structure constant, $\beta=v/c$ is the velocity, $\epsilon(\omega)$ is the complex permittivity of the material, and the $\pm$ terms are positive for particles entering the material and negative for those entering the vacuum.  The angle integrated rate and spectrum of radiation produced is only slightly modified for charged particles away from normal incidence.  Using the measured optical permittivities of silicon at room temperature \cite{Optcal.Const, Optcal.Const2} and copper at 78~K \cite{PhysRevB.11.1315} from the silicon bandgap (1.1 eV) up to a data cutoff $\omega_c \approx 6$~eV.  At energies larger than $\omega_c$, a Druide model continuation is used 
\begin{equation}
\epsilon(\omega) = 1 + \frac{\omega_c^2}{\omega^2}(\epsilon(\omega_c)-1)
\end{equation}

A frequency cutoff occurs where radiated photons are only produced for $\omega < \gamma\omega_p$ where $\gamma$ is the Lorentz boost of the charged particle and $\omega_p$ is the plasmon frequency of the material where $|\epsilon{\omega_p}-1|=1$.  Using the model above, $\omega_p$ is 14.7~eV in silicon and 11.1~eV in copper, within 10\% of their measured values.

The reflection $R_i$ and absorbtion $T_i$ coefficients for photons at normal incidence at the surface of material $i$ are defined by
\begin{equation}
R=\left| \frac{1-\sqrt{\epsilon}}{1+\sqrt{\epsilon}}\right|^2, \qquad T = 1-R
\end{equation}
and can be used to calculate an approximate photon absorbtion efficiency $\eta_i$ in a silicon detector surrounded by copper for photons produced at the surface of material $i$
\begin{gather}
\eta_{\text{Cu}}=\frac{T_{\text{Si}}}{T_{\text{Si}}+R_{\text{Si}}T_{\text{Cu}}}\\
\eta_{\text{Si}}=\frac{R_{\text{Cu}}T_{\text{Si}}}{T_{\text{Cu}}+R_{\text{Cu}}T_{\text{Si}}}\\
\end{gather}
These factors are applied to the spectra of Figure \ref{fig:energy}.

\bibliography{ref}

\end{document}